\documentclass[]{aastex62}
 
\usepackage{natbib}
\newcommand{\HI}{H{\sc ~i}}

\submitjournal{ApJL}

\shorttitle{Impact of Neutral IGM on Ly$\alpha$ Intensity Mapping}
\shortauthors{Visbal and McQuinn}

\begin{document}

\title{The Impact of Neutral Intergalactic Gas on Lyman-$\alpha$ Intensity Mapping During Reionization}

\email{evisbal@flatironinstitute.org}

\author{Eli Visbal}
\affil{Center for Computational Astrophysics, Flatiron Institute, 162 5th Ave, New York, NY, 10003, USA}

\author{Matthew McQuinn}
\affiliation{University of Washington, Department of Astronomy, 3910 15th Ave NE, Seattle, WA, 98195, USA}

\begin{abstract} 
We present the first simulations of the high-redshift Ly$\alpha$ intensity field that account for scattering in the intergalactic medium (IGM). Using a 3D Monte Carlo radiative transfer code, we find that Ly$\alpha$ scattering smooths spatial fluctuations in the Ly$\alpha$ intensity on small scales and that the spatial dependence of this smoothing depends strongly on the mean neutral fraction of the IGM.  Our simulations find a strong effect of reionization on $k=0.1-1~{\rm Mpc^{-1}}$, with $P_{\rm Ly\alpha}\propto k^{-1.75}$ for $\bar{x}_{\rm HI} = 0.63$ and $P_{\rm Ly\alpha} \propto k^{-2.2}$ for $\bar{x}_{\rm HI} = 0.86$ in contrast to  $P_{\rm Ly\alpha}\propto k^{-1.5}$ after reionization. At wavenumbers of $k>1 ~ {\rm Mpc^{-1}}$, we find that the signal is sensitive to the emergent Ly$\alpha$ line profiles from galaxies. We also demonstrate that the cross-correlation between a Ly$\alpha$ intensity map and a future galaxy redshift survey could be detected on large scales by an instrument similar to \emph{SPHEREx}, and over a wide range of scales by a hypothetical intensity mapping instrument in the vein of \emph{CDIM}.
\end{abstract}


\section{Introduction}
When and how the Epoch of Reionization (EoR) occurred is one of the biggest unsolved problems in Cosmology \citep[for a recent review see][]{mcquinn16}.  A well-established observable of this epoch is damping wing absorption in the spectrum of galaxies and quasars, which is sensitive to neutral intergalactic gas that lies tens of megaparsecs in the foreground \citep{Miralda-Escude98}.  Such absorption is the leading explanation for the rapid decline in the number of Lyman-$\alpha$ emitting galaxies above $z\approx 6.6$ \citep{iye06, schenker12, ono12}.  Indeed, many models find that this requires fast evolution in the neutral fraction over this redshift interval \citep{choudhury14, mesinger15}.

There should also be a diffuse \HI\ Ly$\alpha$ component that owes to re-emissions following IGM absorptions; the feasibility of mapping this diffuse intensity is our focus.  Previous work suggests that detecting the total diffuse Ly$\alpha$ flux is possible with the intensity mapping sattelites \emph{SPHEREx} and \emph{CDIM} \citep{2013ApJ...763..132S, 2014ApJ...786..111P, 2016MNRAS.455..725C, 2014arXiv1412.4872D, 2016arXiv160205178C}.  However, previous work only considered the direct diffuse emission from the sum total of all galaxies, ignoring scattering from a neutral IGM that should imprint the structure of reionization on this signal.

This Letter presents calculations of the diffuse Ly$\alpha$ intensity field from the EoR.  In Section 2 and 3, we describe these calculations. Section 4 presents simulated Ly$\alpha$ intensity maps.  Section 5 concludes with a brief discussion of our results and the major takeaways.  Throughout, we assume a $\Lambda {\rm CDM}$ cosmology with parameters consistent with those reported by the \cite{2014A&A...571A..16P}: $\Omega_{\rm m}=0.32$, $\Omega_{\Lambda}=0.68$, $\Omega_{\rm b}=0.049$, $h=0.67$, $\sigma_8=0.83$, and $n_{\rm s}=0.96$. All cosmological distances are given in comoving units.

\section{Reionization Simulations}
We use the 21cmFAST code to generate 200 Mpc cubic realizations of reionization at $z=7$ with a resolution of $256^3$ voxels, outputting the IGM density, the IGM neutral fraction, and the halo mass density field \citep{2011MNRAS.411..955M}.   The density and velocity fields in 21cmFAST are computed with the Zeldovich approximation; the halo positions and ionization fields are calculated using excursion set models.  Excursion set algorithms for reionization successfully reproduce the morphology of reionization in full radiative transfer simulations \citep{2007ApJ...654...12Z}.   There are three main astrophysical parameters that determine the properties of reionization in 21cmFAST: the minimum virial temperature of halos hosting ionizing sources, $T_{\rm min}$, the maximum radius of individual HII regions, $R_{\rm min}$, and the ionizing efficiency, $\zeta$, which encodes quantities such as the star formation efficiency and the escape fraction of ionizing photons  \citep[for details see][]{2011MNRAS.411..955M}. For all runs, we set $T_{\rm min}=5\times10^4~\rm{K}$ (corresponding to a minimum halo mass of $M_{\rm halo, min}=1.5\times10^{9}~M_{\odot}$) and $R_{\rm max}=15~{\rm Mpc}$.  While sources down to $M_{\rm halo,min} \approx 10^{8}~M_{\odot}$ can host substantial star formation, feedback likely makes these halos less efficient \citep{2017MNRAS.472.1576F};  our choice of $M_{\rm halo,min}$ accounts for this inefficiency within the parametrization of 21cmFAST.    Furthermore, $R_{\rm max}$ is set by the mean free path of ionizing photons, which current models suggest is less than $15~{\rm Mpc}$ at $z=7$ \citep{worseck14, daloisio16}.\footnote{These models extrapolate mean free path observations at $z\sim 5$ to higher redshifts.}  Smaller $R_{\rm max}$ or $M_{\rm halo,min}$ would increase the sizes of the effects we investigate. With these parameters fixed, we simulate three different ionizing efficiencies, $\zeta=10$, 15, and 20, to have three different ionized fractions to compare.   

 21cmFAST outputs a neutral fraction field that is either equal to 0 or 1.  This binarity does not account for the small residual hydrogen within ionized regions, which can scatter Ly$\alpha$ photons that redshift through line center.  We set the neutral fraction inside ionized regions to $x_{\rm HI}=10^{-4}$ by hand, with the value chosen to yield a significant line center optical depth (mimicking the saturation seen in the highest redshift Ly$\alpha$ forest).  However, varying this value a factor of a few has no impact on our results.  We do not use 21cmFAST to compute temperatures of the gas and instead for simplicity assume that neutral regions have a uniform temperature of $T=10^3~{\rm K}$ and that ionized regions have a temperature of $T=10^4~{\rm K}$. These temperatures are what could be expected due to X-ray heating in the neutral IGM \citep[e.g.][]{2007MNRAS.376.1680P} and heating associated with reionization in the ionized IGM. We find that setting the entire box to $T=10^3~{\rm K}$ does not change the Ly$\alpha$ power spectrum by more that 10 percent on the spatial scales examined in this Letter.

We assume that Ly$\alpha$ photons are produced by hydrogen recombinations in the interstellar medium of galaxies in halos larger than $T_{\rm min}$. The total Ly$\alpha$ luminosity of a galaxy is related to the star formation rate, $\dot{M}_*$, by
\begin{equation}
\label{luminosity}
L_{\rm gal}=2\times10^{42}\left(1-f_{\rm esc}\right)\frac{\dot{M}_*}{M_\odot~{\rm yr}^{-1}}~{\rm erg~s^{-1}},
\end{equation}
where $f_{\rm esc}$ is the fraction of ionizing photons that escape into the IGM.\footnote{In the absence of dust absorptions, Ly$\alpha$ photons originating from IGM recombinations should be smaller by $f_{\rm esc}/(1-f_{\rm esc})$.} This equation assumes no dust absorption so that every ionization results in $0.6$ Ly$\alpha$ photons, and a Salpeter initial mass function (IMF; \citealt{2003A&A...397..527S}) over a mass range of $1-100 M_\odot$ with metallicity $Z=0.04$; other empirically-motivated PopII IMFs result in factor of $\sim 2$ differences.   Our model assumes SFR proportional to halo mass for all halos with $T>T_{\rm min}$, where the global star formation rate density is normalized to $\rho_\star=0.015~M_\odot~ {\rm yr^{-1}~Mpc^{-3}}$. This value is similar to $\rho_\star = 0.02~M_\odot~{\rm yr^{-1}~Mpc^{-3}}$ measured at $z\approx 6.8$ in \cite{2015ApJ...803...34B}.  Our predicted intensities scale linearly with this overall normalization.  Since we always assume the same SFR prescription, modifying $\zeta$ to change the mean neutral fraction is equivalent to changing $f_{\rm esc}$.  However, we use $f_{\rm esc}=0.1$ in Eq.~\ref{luminosity} to fix the total luminosity (instead of allowing it to change by $5\%$ across our simulations), permitting a more straightforward comparison. Unless otherwise noted, we assume that the emergent Ly$\alpha$ spectrum escaping from the interstellar medium of each star forming galaxy is a Gaussian with a $1-\sigma$ width of $100~{\rm km~s^{-1}}$.  Lastly, to simplify the calculation, we take ten percent of halos to be active Ly$\alpha$ emitters at a given time, making active emitters ten times more luminous to preserve the global emissivity; we comment when this assumption affects our results.

\section{Ly$\alpha$ Radiative Transfer}
To compute the impact of Ly$\alpha$ scattering on simulated intensity maps, we have developed a parallelized 3D Monte Carlo Ly$\alpha$ radiative transfer code, similar to the one described in \cite{2010ApJ...725..633F} \citep[see also][]{2002ApJ...578...33Z, 2005ApJ...628...61C, 2006ApJ...649...14D, 2007ApJ...657L..69L}. For a detailed description see Appendix C in \citet{2010ApJ...725..633F}.  To generate a distant-observer image, our calculations follow the paths of Ly$\alpha$ photons packets as they scatter throughout the simulated neutral gas field.  We follow these packets for spatial steps that are much smaller than our simulation resolution, with specifications chosen for convergence. We also utilize the ``accelerated scheme'' described in Appendix C of \cite{2010ApJ...725..633F} with $x_{\rm crit}=5.0$, which greatly speeds up our calculation without significantly impacting the results.  We performed a series of tests on our code including the static sphere, redistribution function, and inflowing/outflowing spheres, described in Appendix C of \cite{2010ApJ...725..633F}, and find good agreement with the established solutions.   With regard to the calculations in this paper, convergence is achieved following $3\times10^6$ photon packets and for the spatial resolution of our 21cmFAST realizations. 

\section{Results}
\begin{figure*}
\includegraphics[clip=false,keepaspectratio=true, width = 185mm]{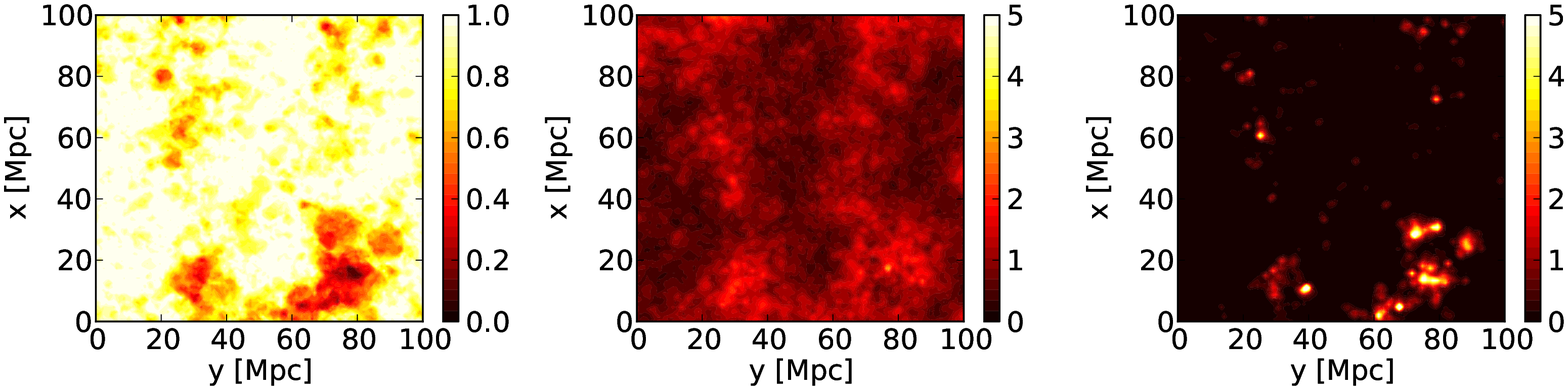}
\includegraphics[clip=false,keepaspectratio=true, width = 185mm]{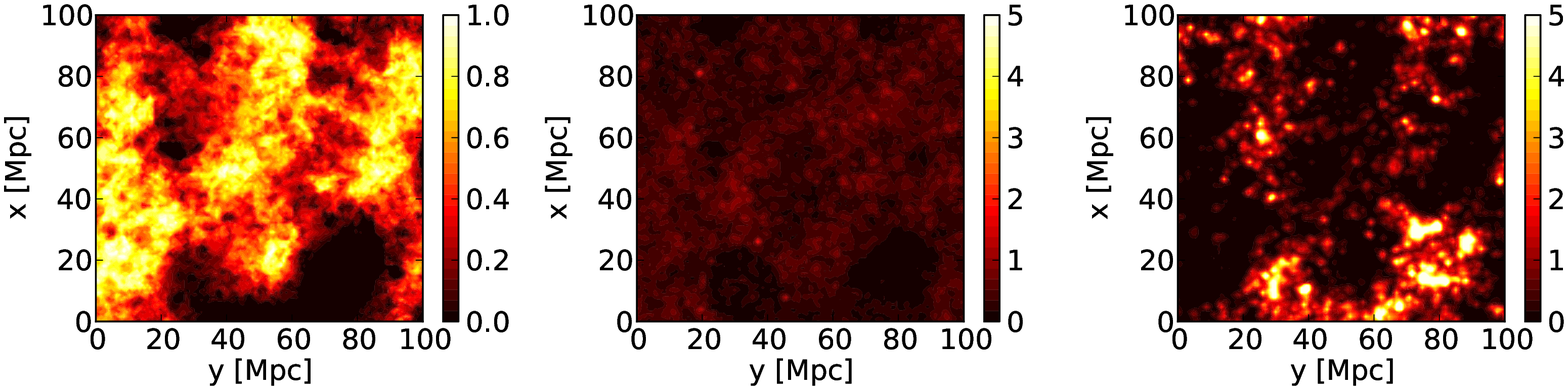}
\caption{\label{slice} The neutral fractions (left column) and corresponding Ly$\alpha$ intensity maps (center and right columns) at $z=7$ and projected over $15$ Mpc for $\bar{x}_{\rm HI}=0.86$ (top row) and $\bar{x}_{\rm HI}=0.36$ (bottom row). The intensity maps have been split into two components: photons with last scatter far from line center ($x>4$, where $x\equiv h|\nu-\nu_\alpha|/k_bT$; center column) and photons with last scatter close to line center with $x\le4$ or that do not scatter at all (right column).  For the calculations in the top row (bottom), 87\% (34\%) of the Ly$\alpha$ photons scatter with $x>4$.  The center column shows radiation that last scattered in the highly neutral IGM, and the right column shows primarily radiation that last scattered from the residual neutral gas within ionized bubbles or radiation that made it direct from the source, not scattering in the IGM.  The Ly$\alpha$ intensity scale shown in the color bars is in units of the mean surface brightness.  
}
\end{figure*}

Here we present our simulated Ly$\alpha$ intensity maps. These maps all have the same underlying dark matter density distribution, but different reionization histories (created by varying the ionizing efficiency parameter in 21cmFAST).  In Figure \ref{slice}, we plot projections of the neutral fraction and Ly$\alpha$ intensity from the $z=7$ snapshot of two of our simulations that have mean neutral fractions of $\bar{x}_{\rm HI}=0.86$ (top row) and 0.36 (bottom row). We separate the Ly$\alpha$ intensity into components with frequencies that are close ($x\le4$) and far ($x>4$) from line center at the final scatter, where $x\equiv h|\nu -\nu_\alpha|/k_bT$ and $\nu_\alpha$ is the frequency of Ly$\alpha$. The component close to line center includes photons that do not scatter at all or that scatter while redshifting through line center within ionized regions. The component far from line center comes mainly from photons that scatter off of highly neutral gas in the IGM and need to redshift into the wing of the Ly$\alpha$ line to escape. The $\bar{x}_{\rm HI}=0.86$ map (top row) has a significantly larger neutral IGM scattered component than the $\bar{x}_{\rm HI}=0.36$.  The enhanced scattering in the former occurs because the ionized bubbles in the IGM are smaller, leading to Ly$\alpha$ photons being less redshifted by the time they reach the edge of the bubbles. Note that for  $\bar{x}_{\rm HI}=0.86$, 0.63, and 0.36, a fraction 0.94, 0.86, and 0.76 of the Ly$\alpha$ photons scatter at least one time, respectively. For the $\bar{x}_{\rm HI}=0.36\;(0.86)$ case, a fraction of $0.34\;(0.87)$ photons scatter in the highly neutral IGM with $x>4$.

\begin{figure}
 \epsscale{0.6}
\plotone{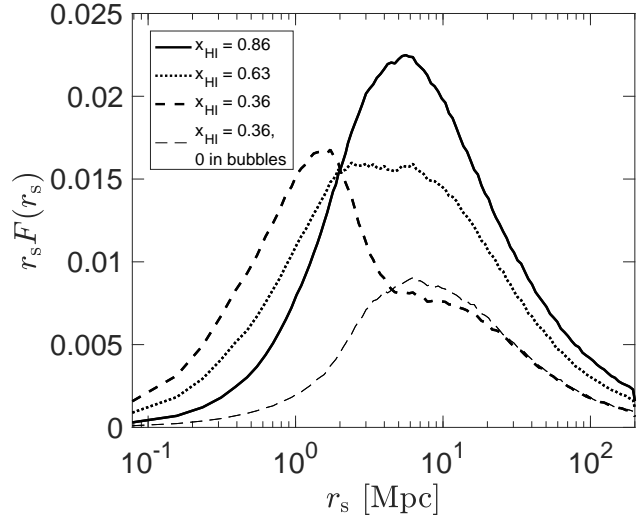}
\caption{\label{sep_hist} Arbitrarily normalized distribution of distances from the source at which the observed photon last scatters, $r_{\rm s}$. We show results for different neutral fractions and for an unphysical case in which the residual neutral fraction inside ionized bubbles is set to zero.
}
\end{figure}

Figure \ref{sep_hist} shows the distribution of distances from the source at which the observed photon last scatters, $F(r_s)$. For the $\bar{x}_{\rm HI}=0.86$ case, $r_s F(r_s)$ has full-width-half-maximum that spans a decade and peaks at $r\approx5~{\rm Mpc}$, slightly smaller than the $\approx 8~$Mpc ionized bubble in a neutral IGM for which a Ly$\alpha$ photon emitted at line (and bubble) center has optical depth unity.  (Another useful limit to note is that of a fully neutral IGM around sources, which yields the $\sim 1~{\rm Mpc}$ in extent \citet{loebrybicki} Ly$\alpha$ `halos'.)  For the case of $\bar{x}_{\rm HI}=0.36$, there is again a peak at $r\approx5~{\rm Mpc}$ if the residual neutral fraction is not included inside ionized bubbles. However, when this residual \HI\ is (realistically) included, there is an additional peak at $r\approx 1~{\rm Mpc}$ that obscures the $r\approx5~{\rm Mpc}$ peak, which primarily arises from photons emitted on the blue-side and that last scatter as they redshift through line center within ionized regions. 

\begin{figure}
\plottwo{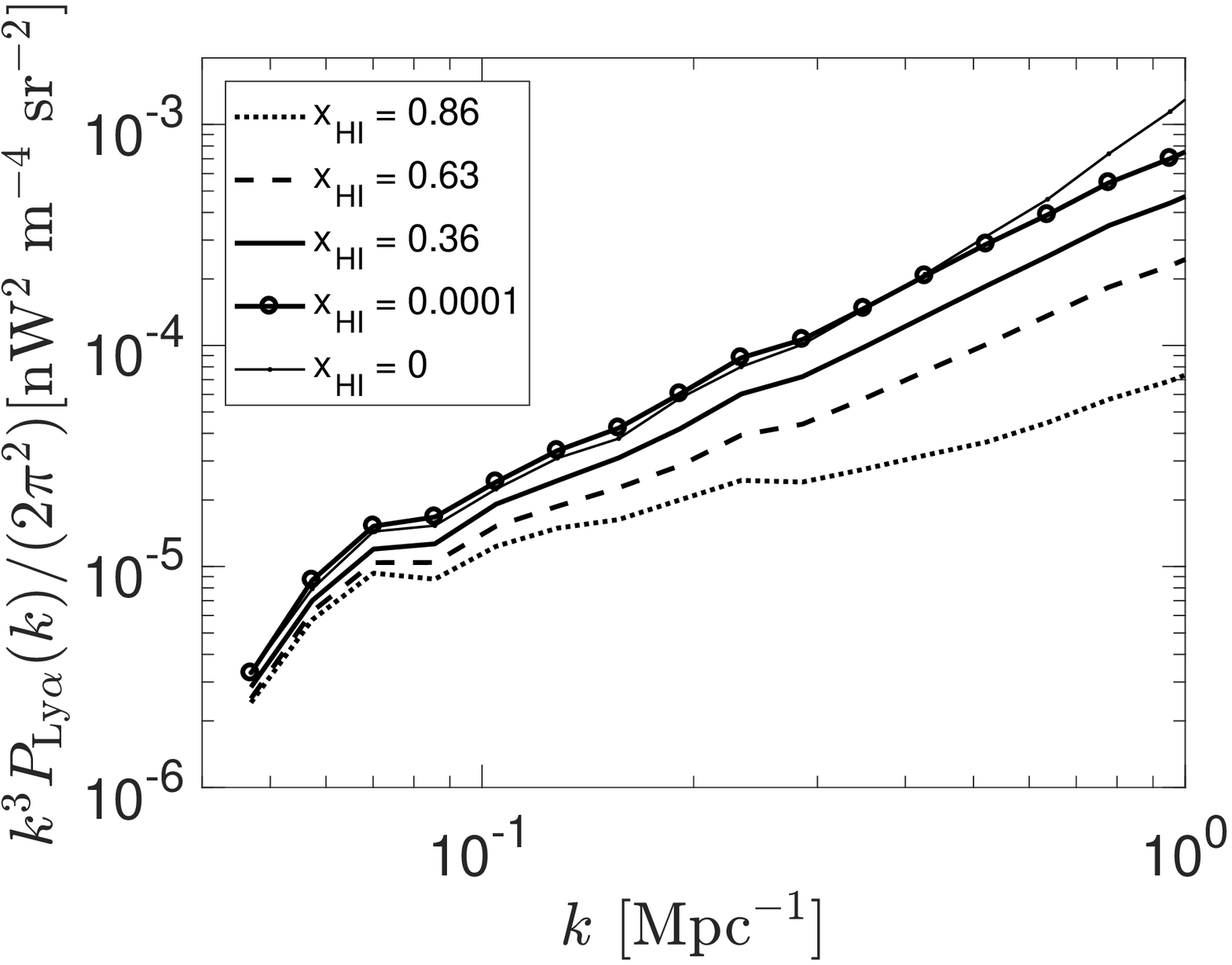}{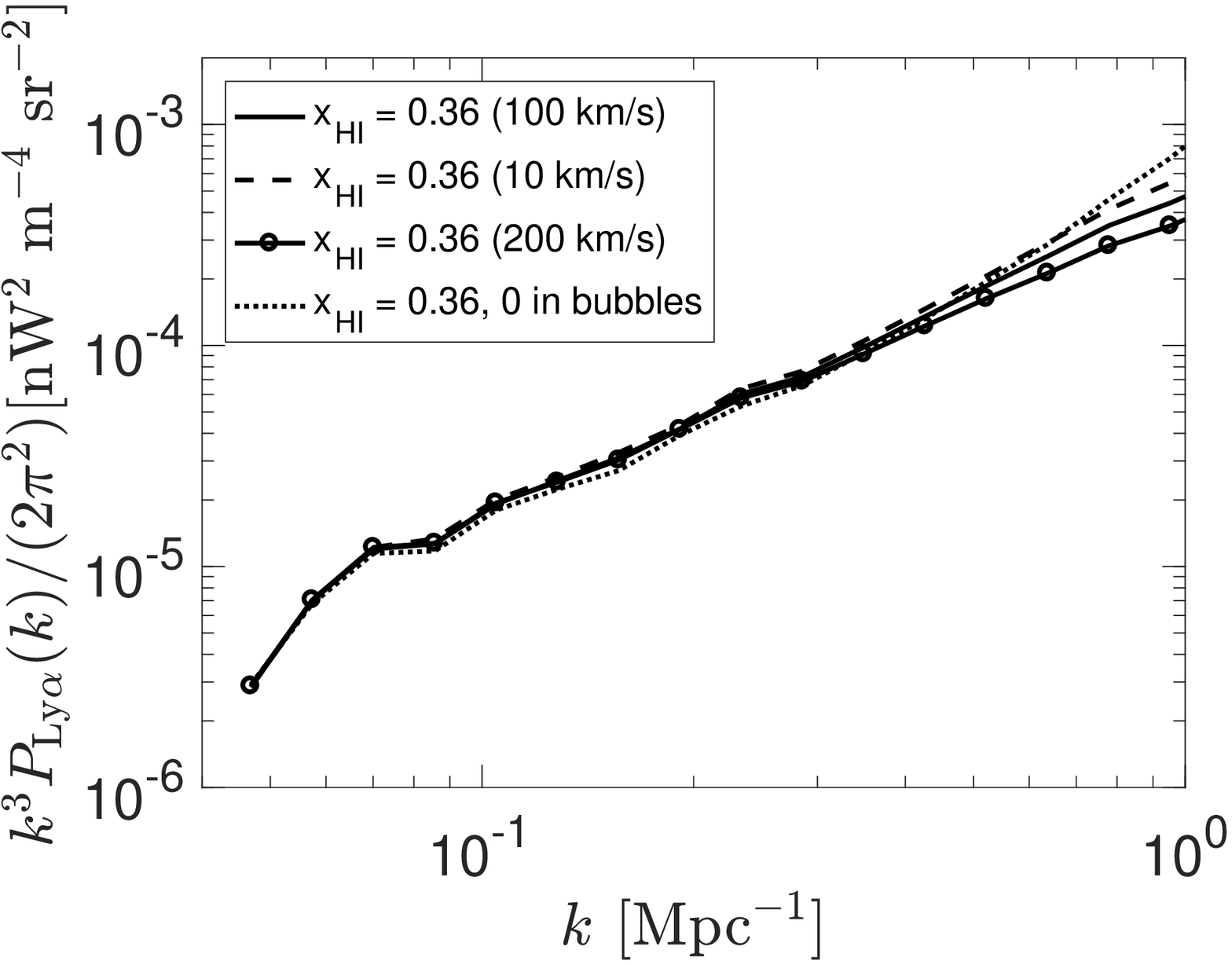}
\caption{\label{power}  Power spectra of our Ly$\alpha$ intensity maps at $z=7$. In the left panel, we show the dependence on the mean neutral fraction of the IGM in our fiducial model. Different neutral fractions are obtained by varying the ionizing efficiency in 21cmFAST for the same realization of the density field. For the $x_{\rm HI}= 10^{-4}$ case, the neutral fraction is uniform. Note that the $x_{\rm HI}=0$ case is unphysical and is what would be computed if radiative transfer were ignored. Ly$\alpha$ scattering in the neutral IGM leads to a scale-dependent suppression in power that depends on the ionization state. In the right panel, we investigate our radiative transfer assumptions. We vary the emergent Ly$\alpha$ line profiles from galaxies (assumed to be Gaussian with $1-\sigma$ width of 100 km/s in the fiducial case) and the residual neutral fraction inside ionized bubbles (changed the fiducial $x_{\rm HI}=10^{-4}$ to $x_{\rm HI}=0$). These assumptions only affect our predictions on small spatial scales.  These models normalize to $\rho_\star=0.015~M_\odot~ {\rm yr^{-1}~Mpc^{-3}}$; the results scale quadratically with $\rho_\star$.  
 } 
\end{figure}

The IGM scattering damps the small-scale fluctuations in the maps. Figure \ref{power} illustrates this damping in the intensity power spectrum.  The model curves' normalizations scale quadratically with the global star formation rate density and so differences in the overall normalization likely cannot be used to detect reionization.  However, differences in the slope of $P_{\rm Ly\alpha}$ on the linear scales that are shown likely cannot be mimicked by uncertainties in the galaxy formation model.  For our fiducial case between $k=0.1-1~{\rm Mpc^{-1}}$, the power spectrum scales roughly as $P_{\rm Ly\alpha} \propto k^{-1.5}$ for $\bar{x}_{\rm HI} = 10^{-4}$,  $P_{\rm Ly\alpha} \propto k^{-1.55}$ for $\bar{x}_{\rm HI} = 0.36$, $P_{\rm Ly\alpha} \propto k^{-1.75}$ for $\bar{x}_{\rm HI} = 0.63$, and $P_{\rm Ly\alpha} \propto k^{-2.2}$ for $\bar{x}_{\rm HI} = 0.86$.   The most significant slope differences require relatively high neutral fractions, which likely occur at higher redshifts than $z\approx 7$.

One of the main challenges for detecting this signal is contamination due to line emission from foreground galaxies \citep{2014ApJ...785...72G, 2016MNRAS.463.3193C}. The most problematic for $z\sim 7$ Ly$\alpha$ is $z\sim 1$ H$\alpha$, which can dwarf the Ly$\alpha$ signal. One way to mitigate this contamination is to cross correlate with a different probe of the same underlying density field. Here we explore the cross correlation of Ly$\alpha$ intensity maps and a future galaxy redshift survey. We compute the Ly$\alpha$-galaxy cross power spectrum and the corresponding sensitivity for two hypothetical space-based intensity mapping telescopes, one roughly modeled after \emph{CDIM} and the other after \emph{SPHEREx}.  Unfortunately, a suitable high-redshift galaxy sample does not yet exist. We assume the galaxies in our survey have a volume density of $n_{\rm gal } = 10^{-4}~{\rm Mpc}^{-3}$ and a mean bias of $b_{\rm gal} = 3$ \citep[similar to the LAEs measured with \emph{Subaru},][although better redshift determinations would be needed for this population]{2018PASJ...70S..13O}.  The dimensions of our survey are assumed to be $3^\circ\times3^\circ$ on the sky and $\Delta z=0.8$ in redshift along the line of sight. The signal-to-noise of the cross power spectrum scales as $b_{\rm gal}n_{\rm gal }^{-1/2}V^{-1/2}$, where $V$ is the survey volume.

\begin{figure*}
\plottwo{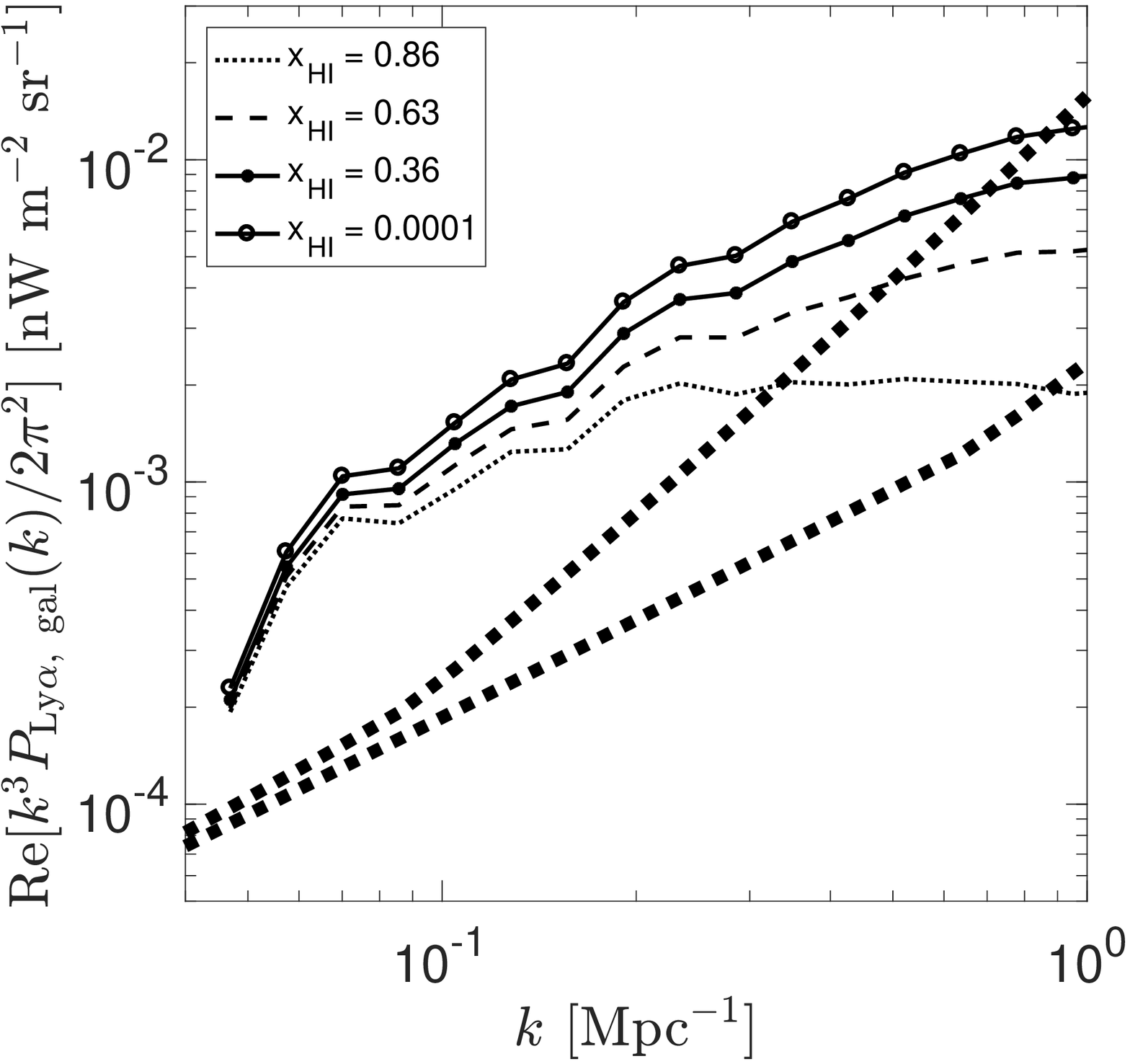}{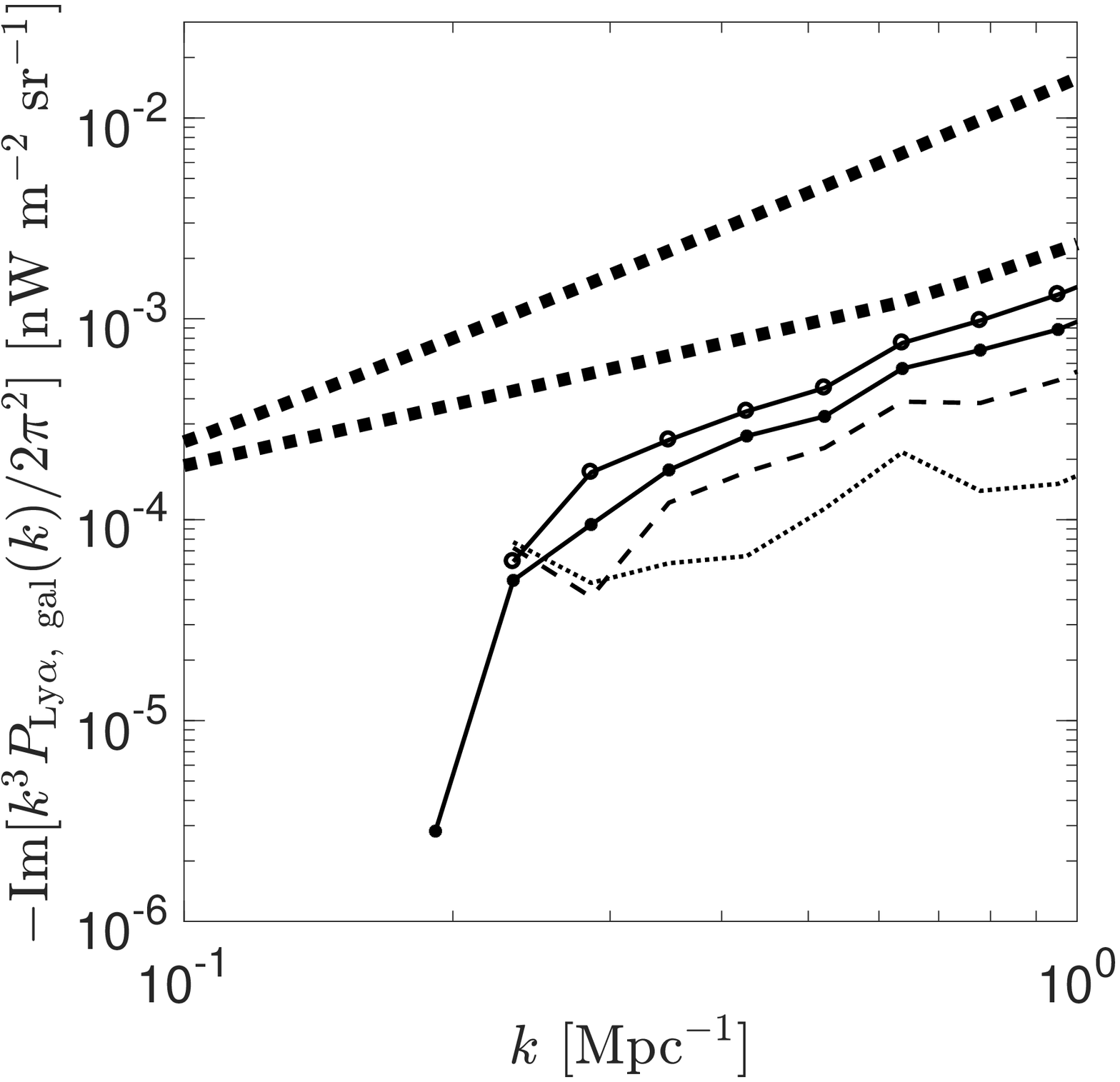}
\caption{\label{cross_power}  The real (left) and imaginary (right) components of the cross power spectrum at $z=7$ between Ly$\alpha$ intensity and the hypothetical galaxy redshift survey described in Section 3 for various neutral fractions.  We show the density-tracing component of this signal (dropping the shot that is likely only relevant at the highest $k$), assuming the galaxies in the redshift survey have bias of $3$.  The upper and lower thick solid dashed lines denote the $1-\sigma$ sensitivity of the hypothetical instruments discussed in Section~3 inspired by \emph{SPHEREx} and \emph{CDIM}, respectively. These sensitivity curves assume band power bins with $\Delta k = k/5$, a field of view of $3^\circ\times3^\circ$, and a depth of $\Delta z=0.8$.} 
\end{figure*}

In Figure \ref{cross_power}, we plot the cross-power spectrum and associated $1-\sigma$ sensitivity for our hypothetical surveys. The cross-power spectrum is estimated by cross-correlating our Ly$\alpha$ intensity maps with the density field from 21cmFAST multiplied by $b_{\rm gal}$. The error on the power spectrum is given by $\delta P_{\rm Ly\alpha,~gal}(k)^2\approx(P_{N}+P_{\rm Ly\alpha}+P_{\rm H\alpha})/(2n_{\rm gal}N_{\rm k})$, where $P_{N}$ is the noise power from our hypothetical Ly$\alpha$ intensity mapping telescopes, $P_{\rm H\alpha}$ is the power from interloping H$\alpha$ sources, and $N_{\rm k}$ is the number of independent modes per $k$-bin. The noise power is calculated using Eqs.~16 and 17 in \cite{2016MNRAS.463.3193C} assuming a telescope diameter of 1.5 m for the \emph{CDIM}-like instrument and 0.2~m for the \emph{SPHEREx}-like instrument, an integration time of $10^5$ s, a zodiacal light background intensity of $\nu I_\nu=500~{\rm nW~m^{-2}~sr^{-1}}$, and an observational efficiency of detecting a photon accounting for losses in the instrument of $\epsilon = 0.5$. We assume spectral resolutions of $R=300$ and $R=40$ for our \emph{CDIM}-like and \emph{SPHEREx}-like instruments, respectively.

The power from foreground H$\alpha$ remaining after masking out the brightest sources is taken to be $P_{\rm H\alpha}=0.04\times({\rm{Mpc}^{-1}}/k)~{\rm{nW}~m^{-2}~sr^{-1}Mpc^3}$. This is roughly the same as computed by \cite{2014ApJ...786..111P} for $z=6$ and a limiting flux above which H$\alpha$ sources can be removed of $10^{-17}~{\rm{erg}~s^{-1}~cm^{-2}}$ (see their Figure 13). This flux cut corresponds to an r-band AB magnitude of $m_r\approx26.5$, which will be observable over large areas with telescopes such as the Hyper Suprime-Cam \citep{2014ApJ...786..111P}. We note that the errors in cross power are dominated by the H$\alpha$ interlopers on large scales for the \emph{SPHEREx}-like instrument and on all scales shown for the \emph{CDIM}-like instrument. Reducing the size of the \emph{CDIM}-like instrument to 0.5 m has very little impact on the results (for a survey this small). We also note that if the flux limit to remove H$\alpha$ sources were an order of magnitude higher (corresponding to $m_r \approx 24$), the unmasked $P_{\rm H\alpha}$ would increase by roughly a factor 30 \citep{2014ApJ...786..111P}. For the \emph{CDIM}-like instrument this increases the errors on the cross power spectrum by $\sim$5.5 for all wavenumbers shown in Figure \ref{cross_power} (since this is the dominant source of error). For the \emph{SPHEREx}-like instrument, this increases the error by a factor of $\sim$5 at $k\sim0.1~{\rm Mpc}^{-1}$ and by a factor of $\sim$2   at $k\sim1.0~{\rm Mpc}^{-1}$. This pessimistic masking criterion would make it challenging to observe the shape of the power spectrum with either instrument.  As in the case of the Ly$\alpha$ power spectrum, the shape of the Ly$\alpha$-galaxy cross power spectrum also depends sensitively on the mean neutral fractions of our reionization simulations. At $k=1~{\rm Mpc}^{-1}$, this results in the cross power spectrum being 1.4, 2.4, and 6.7 times lower for $\bar{x}_{\rm HI}=0.36$, 0.63, and 0.86, respectively, than for $\bar{x}_{\rm HI}=10^{-4}$.

We have investigated two potentially smoking gun signatures for diffuse IGM emission:  An imaginary cross power, and non-isotropy in $\mathbf{k}$. There is $\sim 10~{\rm Mpc}$ (comoving) spatial offset between observed galaxies and Ly$\alpha$ intensity maps along the line of sight due to Ly$\alpha$ photons needing to be redshifted approximately $\sim 1000~{ \rm km ~ s^{-1}}$ to escape to the observer. This leads to a non-zero imaginary component of the cross power spectrum, $\langle \tilde{I}_{\rm \nu, Ly\alpha}(k)  \times \tilde{\delta}_{\rm gal}(k)^* \rangle $, which is shown along with the real component in Figure \ref{cross_power}. However, its low amplitude will make it more challenging to observe. Additionally, because large ionized structures oriented along the line of sight enhance transmission, modes oriented
along the line-of-sight (which lead to smaller ionized structures in this direction) have suppressed power relative to modes perpendicular to the line of sight (which lead to larger ones). We find that the anisotropy in $P_{\rm Ly\alpha}(\mathbf{k})$  is only a large effect for high mean neutral fraction and leave a detailed analysis for future work.

Our choice of duty cycle, $\epsilon_{\rm duty} = 0.1$, has no impact on the cross-power spectrum in our calculation (but it does impact high-$k$ in the auto, increasing $P_{\rm Ly\alpha}(k)$ by a factor of $\sim$2 at $k\sim1~{\rm  Mpc^{-1}}$). We note that in reality there is likely to be some cross shot power between the observed galaxies and Ly$\alpha$ intensity map, that will enhance the signal at high wavenumbers (with the neglected term scaling as $\sim k^3$ in $k^3 P_{\rm Ly\alpha, gal}$).

\section{Summary and Conclusions}
We have simulated the Ly$\alpha$ intensity field from the EoR, including for the first time scattering by neutral intergalactic gas.  We find that the scattering of Ly$\alpha$ photons in the IGM produces a scale-dependent suppression of spatial fluctuations on linear scales that likely cannot be mimicked by other processes, with the magnitude of this suppression dependent on the mean IGM neutral fraction.  We find that using a Ly$\alpha$ intensity map supplied by an instrument similar to \emph{SPHEREx} and especially \emph{CDIM}, this signal at $z\sim7$ could be detected in cross correlation with a future galaxy redshift survey.  In future work, we intend to investigate the dependences of this signal on different reionization scenarios, as well as the effect of additional processes that could enhance the EoR signal (such as Ly$\alpha$ cooling in the ionizing fronts of expanding HII regions and from galaxy continuum photons being redshifted into Lyman-series lines).

\section*{Acknowledgements}
The Flatiron Institute is supported by the Simons Foundation. The simulations were carried out on the Flatiron Institute supercomputer, Rusty. MM was supported by NASA ATP award NNX17AH68G and by the Alfred P. Sloan foundation.

\bibliography{lya_letter}

\end{document}